\documentclass[showpacs,superscriptaddress,preprint]{revtex4-1}

\usepackage{graphicx}
\usepackage{amsmath}
\usepackage{amssymb}
\usepackage{dcolumn}
\usepackage{bm}
\usepackage{color}

\usepackage[utf8]{inputenc}

\allowdisplaybreaks[1]

\newcommand{\bs}[1]{\boldsymbol{#1}}
\newcommand{\dom}{d} 
\newcommand{\cell}{l} 

\usepackage[colorlinks]{hyperref}
\hypersetup{citecolor=blue,linkcolor=blue}
 
\usepackage[capitalise]{cleveref}
\crefname{equation}{Eq.\!}{Eqs.\!}
\crefname{figure}{Fig.\!}{Figs.\!}
\crefname{chapter}{Chap.\!}{Chaps.\!}
\crefname{section}{Sec.\!}{Secs.\!}
\crefname{appsec}{App.\!}{Apps.\!}

\usepackage[a4paper, margin=.75in]{geometry}

\begin{document}

\title{Invariant current approach to wave propagation \\ in locally symmetric structures}

\author{V.\,E. Zampetakis}
\affiliation{International Baccalaureate Department, Geitonas School, 16602 Vari Attikis, Greece}

\author{M.\,K. Diakonou}
\affiliation{Model Lyceum Evangeliki School of Smyrna, 17123 Nea Smirni, Greece}

\author{C.\,V. Morfonios}
\affiliation{Zentrum f\"ur Optische Quantentechnologien, Universit\"{a}t Hamburg, 22761 Hamburg, Germany}

\author{P.\,A. Kalozoumis}
\affiliation{Department of Physics, University of Athens, 15771 Athens, Greece}

\author{F.\,K. Diakonos}
\affiliation{Department of Physics, University of Athens, 15771 Athens, Greece}

\author{P. Schmelcher}
\affiliation{Zentrum f\"ur Optische Quantentechnologien, Universit\"{a}t Hamburg, 22761 Hamburg, Germany}
\affiliation{The Hamburg Centre for Ultrafast Imaging, Universit\"{a}t Hamburg, 22761 Hamburg, Germany}

\date{\today}

\begin{abstract}

A theory for wave mechanical systems with local inversion and translation symmetries is developed employing the two-dimensional solution space of the stationary Schr\"odinger equation.
The local symmetries of the potential are encoded into corresponding local basis vectors in terms of symmetry-induced two-point invariant currents which map the basis amplitudes between symmetry-related points.
A universal wavefunction structure in locally symmetric potentials is revealed, independently of the physical boundary conditions, by using special local bases which are adapted to the existing local symmetries.
The local symmetry bases enable efficient computation of spatially resolved wave amplitudes in systems with arbitrary combinations of local inversion and translation symmetries.
The approach opens the perspective of a flexible analysis and control of wave localization in structurally complex systems.

\end{abstract}


\maketitle

\section{Introduction}

Symmetries play an essential role for the structure and predictions of modern physical theories by their generic relation to constants of motion. 
In classical dynamics, continuous symmetries lead to the conservation of associated currents following from Noether's theorem \cite{Neuenschwander2010____EmmyNoethers} which has subsequently been generalized in various ways \cite{Rosen1974_AP_82_54_GeneralizedNoethers,Vanderschaft1981_SCL_1_108_SymmetriesConservation,Gordon1984_AP_155_85_EquivalentConserved,Anco1996_JMP_37_2361_DerivationConservation}. 
In a quantum description, the relation between symmetry and conservation laws is extended to discrete symmetries \cite{Neuenschwander1995_AJP_63_489_QuestionNoethers,Hernandez-bermejo1996_AJP_64_840_AnswerQuestion} by the commutation of the corresponding operators with the Hamiltonian, thus yielding a connection to the possible form of stationary eigenstates of a system.
In particular, states of definite parity in inversion-symmetric systems and conserved quasimomenta in structures with discrete translation invariance (to be referred to as parity and Bloch theorems, respectively) are central to the treatment and understanding of a large class of phenomena in, e.\,g., atoms or crystals.

\begin{figure}
\def\svgwidth{.8\columnwidth}
\begingroup%
  \makeatletter%
    \setlength{\unitlength}{\svgwidth}%
  \begin{picture}(1,0.31666304)%
    \put(0,0){\includegraphics[width=\unitlength]{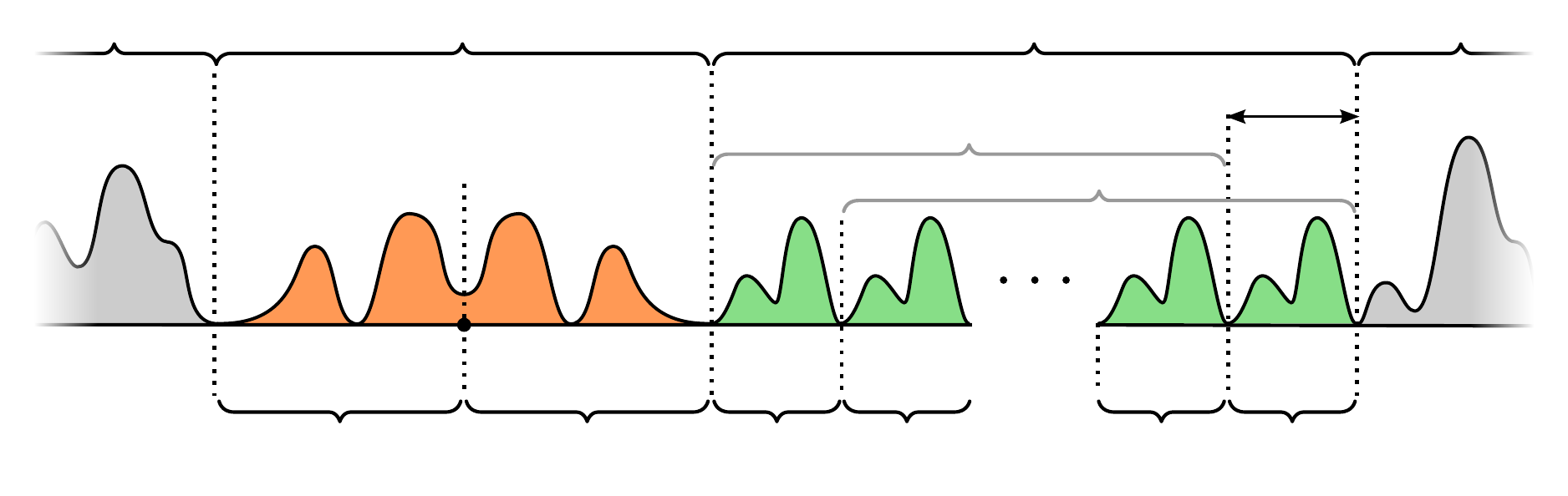}}%
    \put(0.29502398,0.3052595){\color[rgb]{0,0,0}\makebox(0,0)[b]{\smash{$\mathcal{D}_2=\bar{\mathcal{D}}_2$}}}%
    \put(0.66039207,0.3052595){\color[rgb]{0,0,0}\makebox(0,0)[b]{\smash{$\mathcal{D}_3$}}}%
    \put(0.6159627,0.24009641){\color[rgb]{0,0,0}\makebox(0,0)[b]{\smash{$\mathcal{D}^\circ_3$}}}%
    \put(0.70185949,0.20455291){\color[rgb]{0,0,0}\makebox(0,0)[b]{\smash{$\bar{\mathcal{D}}^\circ_3$}}}%
    \put(0.21609828,0.00313968){\color[rgb]{0,0,0}\makebox(0,0)[b]{\smash{$\mathcal{C}^{(2)}_1$}}}%
    \put(0.37308209,0.00313968){\color[rgb]{0,0,0}\makebox(0,0)[b]{\smash{$\mathcal{C}^{(2)}_2$}}}%
    \put(0.49452239,0.00313968){\color[rgb]{0,0,0}\makebox(0,0)[b]{\smash{$\mathcal{C}^{(3)}_1$}}}%
    \put(0.57745723,0.00313968){\color[rgb]{0,0,0}\makebox(0,0)[b]{\smash{$\mathcal{C}^{(3)}_2$}}}%
    \put(0.74036496,0.00313968){\color[rgb]{0,0,0}\makebox(0,0)[b]{\smash{$\mathcal{C}^{(3)}_{N_3-1}$}}}%
    \put(0.8232998,0.00313968){\color[rgb]{0,0,0}\makebox(0,0)[b]{\smash{$\mathcal{C}^{(3)}_{N_3}$}}}%
    \put(0.93141129,0.3052595){\color[rgb]{0,0,0}\makebox(0,0)[b]{\smash{$\mathcal{D}_4$}}}%
    \put(0.296161,0.21221199){\color[rgb]{0,0,0}\makebox(0,0)[b]{\smash{$\alpha$}}}%
    \put(0.82181882,0.25194424){\color[rgb]{0,0,0}\makebox(0,0)[b]{\smash{$L$}}}%
    \put(0.1361254,0.08237207){\color[rgb]{0,0,0}\makebox(0,0)[b]{\smash{$x_1$}}}%
    \put(0.45305497,0.08237207){\color[rgb]{0,0,0}\makebox(0,0)[b]{\smash{$x_2$}}}%
    \put(0.86402673,0.08237207){\color[rgb]{0,0,0}\makebox(0,0)[b]{\smash{$x_3$}}}%
    \put(0.07244329,0.3052595){\color[rgb]{0,0,0}\makebox(0,0)[b]{\smash{$\mathcal{D}_1$}}}%
  \end{picture}%
\endgroup%
\caption{ 
Local symmetry decomposition of a potential region into inversion ($\varPi$) or translation ($T$) symmetric domains $\mathcal{D}_\dom$ of size $x_\dom - x_{\dom-1}$ which are further decomposed into $N_\dom$ cells $\mathcal{C}^{(\dom)}_{l=1:N_\dom}$.
The $\varPi$-symmetric domain $\mathcal{D}_2$ (with $N_\dom = 2$ cells) is mapped onto itself under inversion through $\alpha$, while in the $T$-symmetric domain (with $N_\dom \geqslant 2$ cells) $\mathcal{D}_3$ the first $N_d - 1$ cells (denoted by $\mathcal{D}^\circ_3$) are mapped onto the last $N_d - 1$ cells (denoted by $\bar{\mathcal{D}}^\circ_3$) under translation by $L$.
Domains without symmetry (like, e.\,g., $\mathcal{D}_1$ and $\mathcal{D}_4$) may in general also be part of the potential region.
}
\label{fig:ls_domains}
\end{figure}
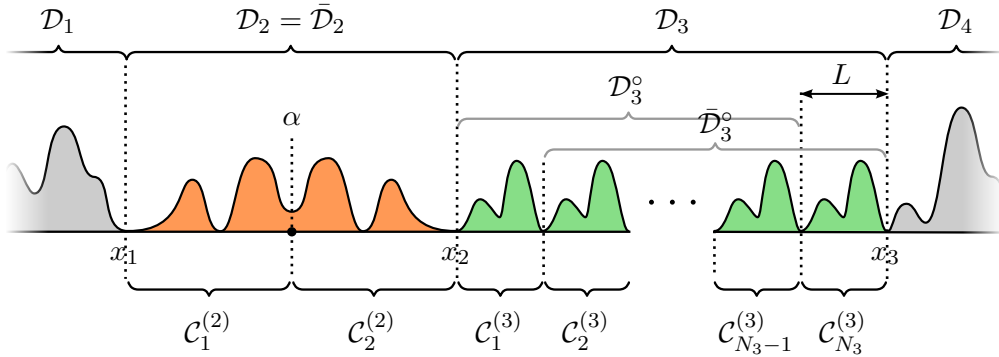

The significance of symmetries is perhaps most appreciated when they are broken \cite{Anderson1972_S_177_393_MoreDifferent}, either explicitly at the level of the equations of motion or spontaneously by the system state itself \cite{Castellani2003_SP___MeaningSymmetry}.
Symmetry breaking is thereby commonly related to emergent effective interactions \cite{Anderson1972_S_177_393_MoreDifferent} or to (ground) state properties \cite{Weinberg1996____QuantumTheory}, a prominent example being the origin of particle mass in the Higgs mechanism \cite{Bernstein1974_RMP_46_7_SpontaneousSymmetry}.
Regarding spatial transformations, the Hamiltonian of a composite system may obey a symmetry only in a subpart of configuration space, in which case the symmetry is broken globally.
This restricted occurrence of a spatial symmetry constitutes a kind of symmetry breaking which is in fact unavoidable due to the finite size of any actual system. 
In reality any symmetry of the effective potential describing a system is indeed restricted to some finite spatial region, while multiple symmetries may occur domain-wise (see \cref{fig:ls_domains} for an illustration of a composite system described by a potential with different
symmetries in different domains). 
Such `local' spatial symmetries \cite{IUC_ODC___LocalSymmetry} may occur inherently in complex systems such as large molecules \cite{Pascal2001_JPCA_105_9040_ConciseSet,Mishra2014_AC_126_13318_StudyingMicrostructure,Domagaa2008_JAC_41_1140_OptimalLocal}, in quasicrystals \cite{Shechtman1984_PRL_53_1951_MetallicPhase,Widom1989_PRL_63_310_Transfer-matrixAnalysis,Morfonios2014_ND_78_71_LocalSymmetry}, or even in partially disordered matter \cite{Wochner2009_PNAS_106_11511_X-rayCross,Kim2013_JAC_46_1331_DeterminationFluctuations}.
They are also often present by design in, e.\,g., multilayered photonic devices \cite{Macia2006_RPP_69_397_RoleAperiodic,Zhukovsky2010_PRA_81_053808_PerfectTransmission,Peng2002_APL_80_3063_Symmetry-inducedPerfect}, semiconductor superlattices \cite{Ferry1997____TransportNanostructures}, acoustic waveguides \cite{Hladky-hennion2013_JAP_113_154901_AcousticWave,Theocharis2014_NJP_16_093017_LimitsSlow} or magnonic systems \cite{Hsueh2011_JOSAB_28_2584_FeaturesPerfect}.
In such artificial setups, broken global symmetry is often required to obtain structures suitable for specific applications. 
A special case are completely locally symmetric (CLS) setups, where the active region is covered exclusively by domains
with local symmetries \cite{Kalozoumis2014_PRL_113_050403_InvariantsBroken}.

Despite their omnipresence, local symmetries and their consequences in wave mechanical systems are largely overlooked when passing from global to no symmetry:
Although it is very common to treat a composite structure in terms of its coupled subparts, their eventual local symmetries are seldom encoded in the description or directly exploited in calculations.
A first step towards an approach addressing local symmetries was taken in Ref.\,\cite{Kalozoumis2013_PRA_87_032113_LocalSymmetries} by defining local inversion operators and relating them to one-dimensional (1D) wave scattering via associated locally, i.\,e. domainwise, invariant quantities.
These symmetry-induced invariants, which have the form of two-point currents, were subsequently used to classify perfect transmission states in terms of their spatial profile \cite{Kalozoumis2013_PRA_88_033857_LocalSymmetries}.
The generic spatial structure of stationary states in the potential regions of local inversion or translation symmetry was recently established \cite{Kalozoumis2014_PRL_113_050403_InvariantsBroken} by a mapping relation 
\begin{equation}
\label{eq:pure_mapping}
 \psi(\bar{x}) = \frac{1}{J} \left[\tilde{Q}\psi(x)-Q\psi^*(x) \right]
\end{equation}
between the wave amplitude $\psi$ at symmetry-related points $x,\bar{x}$, where the complex two-point currents 
\begin{align}
 Q &= \frac{1}{2i} \left[\sigma\psi(x)\psi^{\prime}(\bar{x}) - \psi^{\prime}(x)\psi(\bar{x}) \right], \label{eq:Qpure} \\
 \tilde{Q} &= \frac{1}{2i}\left[\sigma\psi^*(x)\psi^{\prime}(\bar{x}) - \psi^{*{\prime}}(x)\psi(\bar{x}) \right]
\end{align}
are invariant, i.\,e. spatially constant, within the corresponding domain of local symmetry.
We here use the notation $\psi'(\bar{x}) = \left.\frac{d\psi(x)}{dx}\right|_{x=\bar{x}}$
Their values depend on the details of the potential via $\psi$ but are related to the globally invariant current $J$ by
\begin{equation}
 |\tilde{Q}|^2 - |Q|^2 = J^2
\end{equation}
in Hermitian systems.

The above mapping relation generalizes the parity and Bloch theorems to systems where reflection and translation symmetries, respectively, are realized only domain-wise \cite{Kalozoumis2014_PRL_113_050403_InvariantsBroken}.
In fact, $Q$ contains information on how a global symmetry is broken:
it vanishes in the case of global symmetry of both the potential and the boundary conditions, in which case \cref{eq:pure_mapping} can be written as the corresponding well-known eigenvalue problems (see Ref.\,\cite{Kalozoumis2014_PRL_113_050403_InvariantsBroken}). 
$Q$ becomes nonzero and globally constant for asymmetric boundary conditions.
In the case of a local symmetry holding in a certain domain, $Q$ and $\tilde{Q}$ are constant in this domain.
Interestingly, $Q$ remains invariant even in the presence of (locally) symmetric complex potentials (in contrast to the usual current $J$), as verified experimentally for CLS acoustic waveguides \cite{Kalozoumis2015_AP___InvariantCurrents}.
It also proves suitable as an order parameter for globally time-parity-symmetric systems, as shown in Ref.\,\cite{Kalozoumis2014_PRA_90_043809_SystematicPathway}.
Focusing on the wavefunction mapping induced by local discrete symmetries \cite{Kalozoumis2014_PRL_113_050403_InvariantsBroken}, we notice that the applicability of \cref{eq:pure_mapping} depends on the boundary conditions imposed on the stationary state $\psi$, since the mapping coefficients diverge for $J=0$.
While there is always some finite transmission in usual 1D scattering settings, the current typically vanishes for any bound eigenstate (as well as for scattering eigenstates of the inversion operator \cite{Kalozoumis2013_PRA_87_032113_LocalSymmetries}), and therefore
an equally valid symmetry mapping for such cases is desirable.

In the present work we develop a formalism for locally symmetric wave mechanical systems incorporating the above concepts in a form which is independent of the boundary conditions imposed on the physical setup at hand.
Formulated directly in the (two-dimensional) solution space of the stationary Schr\"odinger equation, the theory uniformly encodes local inversion and translation symmetries of the potential into corresponding local basis vectors.
In particular, a construction scheme for a global basis of an arbitrary one-dimensional wave mechanical system is deviced, exploiting the presence of multiple local symmetries in the underlying potential. 
Sets of linearly independent solutions, whose role is central for various types of Sturm-Liouville problems \cite{Khmelnytskaya2014_MMAS_38_1945_EigenvalueProblems,Kravchenko2014_AMC_238_82_ModifiedSpectral}, are hereby utilized to construct generalized two-point currents of mixed basis functions which are spatially constant within domains of local symmetry. 
These invariants establish a basis function mapping between symmetry-related points in each local symmetry domain which incorporates the potential symmetry in the domain basis without explicit reference to the spatial dependence of its components.
The local symmetry bases are then matched to assemble a global basis in terms of arbitrary initial local basis solutions defined in a subspace (cell) of each domain and the associated invariant currents.
Boundary conditions are finally imposed on a physical solution expanded in this global basis.
The introduced procedure is in contrast with the usual approach for solving stationary Schr\"odinger (or Helmholtz) equation in systems with symmetries in restricted domains, where: 
(1) it is necessary to assume that within a domain the basis is the same as that of a system with the respective global symmetry and 
(2) the explicit form of the domain basis is used to construct the overall solution with appropriate matching. 
The introduced formalism of local symmetry bases leads to a threefold main attainment:
(i) The local mapping relations pertaining to the generalized 1D parity and Bloch theorems of Ref.\,\cite{Kalozoumis2014_PRL_113_050403_InvariantsBroken} are extended to arbitrary boundary conditions (i.e. including ones yielding zero current) in terms of the introduced local basis invariants.
(ii) A universal structure of stationary wavefunctions in locally symmetric systems is revealed by their representation in local symmetry-adapted bases which are constructed from arbitrary initial solutions.
(iii) Exploiting the diagonal form of the amplitude mapping matrices in the local symmetry basis (LSB) of each symmetry domain provides an efficient scheme for computing wavefunctions for locally symmetric systems.

The paper is organized as follows.
In \cref{sec:mapping} the symmetry-induced amplitude mapping in terms of two-function, two-point domainwise invariants is derived and the LSB, leading to diagonal mapping matrices, is constructed.
Section \ref{sec:physical_solution} introduces the connection of LSBs of different domains, illustrating the generic structure of the physical solution, and demonstrates its efficient computation with arbitrary boundary conditions with initial input only in single unit cells of each domain.
In \cref{sec:conclusion} we summarize our work and provide concluding remarks.

\section{Locally invariant symmetry mapping}
\label{sec:mapping}

Consider a potential $V$ which is symmetric under the linear coordinate transform 
\begin{equation}
\label{eq:F}
F:x \rightarrow \overline{x} = F(x) = \sigma x + \rho = 
\begin{cases} 2\alpha - x &(F = \varPi) \\ x + L &(F = T) \end{cases}
\end{equation}
within a domain $\mathcal{D} \subseteq \mathbb{R}$, that is, $V(\bar{x}) = V(x)$ under an inversion ($\varPi$) through a point $\alpha$ or a translation ($T$) by a length $L$ (where it is understood that a $T$-symmetry transform applies to $x \in \mathcal{D}^\circ \equiv \{\mbox{all but the last unit cell of } \mathcal{D}\}$; see \cref{fig:ls_domains}).
Any solution $\psi(x)$ to the stationary Schr\"odinger equation for this potential (setting $\hbar = m = 1$),
\begin{equation}
\label{eq:schrodinger} 
H\psi(x) \equiv -\frac{1}{2}\psi''(x) + V(x) \psi(x) = E\psi(x),
\end{equation}
can be expanded in the basis of two linearly independent solutions $\phi_1(x)$, $\phi_2(x)$ of \cref{eq:schrodinger} in $\mathcal{D}$ for a given energy eigenvalue $E$. 
Subtracting $\phi_n(\bar{x})H\phi_m(x)$ from $\phi_m(x)H\phi_n(\bar{x})$, with $m,n \in \{1,2\}$, leads to
\begin{equation}
\label{eq:total_derivative} 
2iq_{m\bar{n}}' \equiv \phi_m(x)\phi_n''(\bar{x}) - \phi_n(\bar{x})\phi_m''(x) = 0
\end{equation}
within $\mathcal{D}$ due to the $F$-symmetry of the potential. 
This means that the `mixed' (i\,e., containing both $\phi_1$ and $\phi_2$) symmetry-induced two-point quantities
\begin{equation} \label{eq:q_mn} 
q_{m\bar{n}} = \frac{1}{2i}  \left[\sigma\phi_m(x)\phi_n^{\prime}(\bar{x}) - \phi_m^{\prime}(x)\phi_n(\bar{x}) \right] 
\end{equation}
are spatially constant within the symmetry domain $\mathcal{D}$ of the potential.
In the same manner an alternative invariant quantity 
\begin{equation} \label{eq:q_mn_tilde} 
\tilde{q}_{m\bar{n}} = \frac{1}{2i}  \left[\sigma\phi_m^*(x)\phi_n^{\prime}(\bar{x}) - \phi^{*\prime}_{m}(x)\phi_n(\bar{x}) \right]
\end{equation}
is obtained, whose translation ($\sigma = 1$) variant for $m=n$ and $\bar{x}=x$ becomes the current 
\begin{equation}
\label{J} 
j_{m}=\frac{1}{2i}  \left[\phi_m^*(x)\phi_m^{\prime}(x)-\phi_m^{*\prime}(x)\phi_m(x) \right]
\end{equation} 
corresponding to the solution $\phi_m$.
The invariants $q_{m\bar{n}}$ and $\tilde{q}_{m\bar{n}}$ thus have the form of mixed nonlocal currents, and for $m=n$, i.\,e. by replacing both $\phi_1(x)$ and $\phi_2(x)$ in \cref{eq:q_mn,eq:q_mn_tilde} with a single solution $\psi(x)$, they reduce to the `pure' (one-function) nonlocal currents $Q$ and $\tilde{Q}$ in \cref{eq:pure_mapping}.
With some algebra it can be shown that the three spatial invariants ${q}_{m\bar{n}}$, $\tilde{q}_{m\bar{n}}$ and $j_m$ are connected via the relation
\begin{equation} \label{eq:eq:Q_new_vsJ} 
|\tilde{q}_{m\bar{n}}|^{2}-|q_{m\bar{n}}|^{2} = j_{m}j_{n},
\end{equation}
which introduces symmetry-induced constraints between the values of a single solution $\psi$ or any pair of solutions $\phi_1, \phi_2$ at $x$ and $\bar{x}$.
Before proceeding, it is worth mentioning that invariants analogous to the form in \cref{eq:q_mn} can be derived also for more general symmetry transformations, as shown in the Appendix.
We here restrict the presentation to the isometry transformations in \cref{eq:F}, which leave the 1D Schr\"odinger equation invariant.

\subsection{Symmetry mapping with mixed currents}
\label{sec:mixed_mapping}

The mixed invariants $q_{m\bar{n}}$ will now be used to construct a general mapping relation between bases in the solution space.
We first write \cref{eq:q_mn} in the matrix form
\begin{align} \label{eq:mixed_mapping_derivation}
 \begin{pmatrix} q_{1\bar{1}} & q_{1\bar{2}} \\ q_{2\bar{1}} & q_{2\bar{2}} \end{pmatrix} = 
 \frac{1}{2i} 
 \begin{pmatrix} \phi_1(x) & \phi_1'(x) \\ \phi_2(x) & \phi_2'(x)  \end{pmatrix} \begin{pmatrix} \sigma & 0 \\ 0 & -1 \end{pmatrix} 
 \begin{pmatrix} \phi_1'(\bar{x}) & \phi_2'(\bar{x}) \\ \phi_1(\bar{x}) & \phi_2(\bar{x})  \end{pmatrix}.
\end{align}
Left-multiplying by the inverse of the first matrix product on the right hand side and then transposing, we can map the solution column vector $\bs{\phi} \equiv \tiny{\begin{pmatrix} \phi_1 \\ \phi_2 \end{pmatrix}}$ between the symmetry-related points $x$ and $\bar{x}$ as
\begin{equation} \label{eq:mixed_mapping}
 \bs{\phi}(\bar{x}) = \bs{Q} \bs{\phi}(x)
\end{equation}
via the (spatially invariant in $\mathcal{D}$) symmetry-mapping matrix
\begin{equation} \label{eq:Q_matrix}
  \bs{Q} = \frac{2i}{w} \begin{pmatrix} -q_{2\bar{1}}  & q_{1\bar{1}} \\  -q_{2\bar{2}} &  q_{1\bar{2}}\end{pmatrix},
\end{equation}
where $w[\phi_1,\phi_2] = \phi_1(x)\phi_2'(x)-\phi_2(x)\phi_1'(x)$ is the Wronskian of the two functions $\phi_1,\phi_2$.
If these are linearly independent solutions of \cref{eq:schrodinger} in an interval $\mathcal{D}$, then $w(x) \neq 0$ $\forall x \in \mathcal{D}$ \cite{Felder2016_MathematicalMethods}.
Note that \cref{eq:mixed_mapping}, with the $q_{m\bar{n}}$ defined in \cref{eq:q_mn}, holds generically for arbitrary potential and for any coordinate transformation $F$.
It is in the presence of $\varPi$- and $T$-symmetry that it becomes a mapping relation with constant (in $\mathcal{D}$) coefficients.
For local $F$-symmetry with an arbitrary smooth transformation $F$, a generalization of the bilinear mapping in \cref{eq:mixed_mapping} also exists, as shown in the Appendix.

It can be shown that the $\bs{Q}$-matrix is unimodular with the determinant $\text{det} \bs{Q} = \sigma$ distinguishing between the case of inversion and translation symmetry.
Relation (\ref{eq:mixed_mapping}) maps any pair of linearly independent solutions from $x$ to the transformed point $\bar{x}$ in a domain via mixed symmetry-induced invariants.
In this sense, it further generalizes the local parity and Bloch theorems of Ref.\,\cite{Kalozoumis2014_PRL_113_050403_InvariantsBroken}, now formulated at the level of the solution space of the Schr\"{o}dinger equation without reference to boundary conditions imposed on physical solutions.
Additionally, \cref{eq:mixed_mapping} indicates that $\phi_1$ and $\phi_2$, although linearly independent, are in general interrelated via local symmetry, in the sense that their values at any symmetry-related points are coupled by the same constant matrix $\bs{Q}$.
A decoupled pair of solutions amounts to a diagonal $\bs{Q}$, as elaborated on in \cref{sec:local_basis}, and is key to deriving an optimal basis for the treatment of stationary wave mechanical problems involving local symmetries, as discussed in \cref{sec:physical_solution}. 

For a physical solution $\psi$ in $\mathcal{D}$ (obeying the appropriate boundary conditions) which is linearly independent from its complex conjugate $\psi^*$, \cref{eq:mixed_mapping} reproduces the pure mapping of \cref{eq:pure_mapping} (and its complex conjugate) if we choose the basis $\bs{\phi}$ to be $\phi_1 = \psi$, $\phi_2 = \psi^*$.
In general, however, $\psi$ and $\psi^*$ are not linearly independent, as is the case for stationary bound states (which can be chosen real, $\psi = \psi^*$) or for stationary scattering eigenstates of the inversion operator $\varPi$ \cite{Kalozoumis2013_PRA_87_032113_LocalSymmetries}.
In these situations the current $J$ vanishes (since it is given by the Wronskian $w[\psi,\psi^*] = 2iJ$, which vanishes if $\psi$, $\psi^*$ are linearly dependent solutions), and the mapping relation of \cref{eq:pure_mapping} cannot be used.
This limitation arises from the fact that \cref{eq:pure_mapping} refers to a physical solution satisfying specific boundary conditions. 
The main advantage of the present approach is that symmetry-induced mapping relations are expressed at the level of the general two-dimensional solution space which is not subject to specific boundary conditions.
Hence, with the basis functions $\phi_1,\phi_2$ in any given domain being linearly independent by assumption (so that $w[\phi_1,\phi_2] \neq 0$), the local mapping (\ref{eq:mixed_mapping}) between symmetry-related points can always be exploited to construct a global basis $\bs{\xi}$, as shown below in \cref{sec:physical_solution}.
Expressed in this global basis $\bs{\xi}$, a physical solution $\psi$ may then have $J=0$ and thus prevent the use of \cref{eq:pure_mapping}, but the local symmetries of the potential are already addressed in the construction of $\bs{\xi}$ using the basis mappings in \cref{eq:mixed_mapping}.
In other words, with the present generalized mapping relation via arbitrary sets of independent domain solutions, the presence of potential symmetry is manifest in any stationary state (also with vanishing current) through the underlying basis:
In any $F$-symmetric domain $\mathcal{D}$, a physical solution $\psi(x) = \bs{a} \cdot \bs{\phi}(x) = a_1 \phi_1(x) + a_2 \phi_2(x)$ of \cref{eq:schrodinger} (with $x \in \mathcal{D}$ and $a_1, a_2$ determined by imposed boundary conditions) at energy $E$ contains the symmetry information through the basis $\phi_{1,2}$ which is mapped between $F$-transformed points via a constant $\bs{Q}$-matrix according to \cref{eq:mixed_mapping}.

\subsection{Local symmetry basis}
\label{sec:local_basis}

Let us now exploit the local $F$-symmetry in the given domain $\mathcal{D}$ to arrive at a decoupled form of the mapping relation in \cref{eq:mixed_mapping}.
To this end, consider the transformation of $\bs{\phi}$ via an invertible constant matrix $\bs{S}$ into the special basis
\begin{equation} \label{eq:S_transform}
 \bs{\chi} 
 = \begin{pmatrix} \chi_+ \\ \chi_- \end{pmatrix}
 = \bs{S} \bs{\phi}
\end{equation}
of the two linearly independent solutions $\chi_+,\chi_-$ in $\mathcal{D}$ which fulfill the mixed conditions
\begin{equation} \label{eq:loc_bas_bound_cond}
 \sigma\chi_\pm(x)\chi^{\prime}_\pm(\bar{x}) = \chi^{\prime}_\pm(x)\chi_\pm(\bar{x})
\end{equation}
for any $x \in \mathcal{D}$, that is, which has $q_{m\bar{m}} = 0$ ($m=+,-$) in $\mathcal{D}$.
In this basis, the invariant matrix performing the symmetry-induced mapping 
\begin{equation} \label{eq:chi_mapping}
 \bs{\chi}(\bar{x}) = \bs{S}\bs{\phi}(\bar{x}) =  \bs{S}\bs{Q}\bs{\phi}(x) = \bs{S}\bs{Q}\bs{S}^{-1}\bs{\chi}(x)
\end{equation}
within $\mathcal{D}$ is diagonal,
\begin{equation} \label{eq:Q_LSbasis}
  \bs{Q}_{\chi} \equiv \bs{S}\bs{Q}\bs{S}^{-1} = \begin{pmatrix} z_+ & 0 \\ 0 & z_- \end{pmatrix}, 
\end{equation}
since the offdiagonal elements $q_{m\bar{m}}$ ($m=+,-$) vanish due to \cref{eq:loc_bas_bound_cond}, with the eigenvalues
\begin{equation} \label{eq:eigenvalues} 
z_\pm = \frac{\text{tr}\bs{Q}}{2}  \pm \sqrt{\left(\frac{\text{tr}\bs{Q}}{2}\right)^2 - \sigma} ~\equiv~ \frac{\text{tr}\bs{Q}}{2}  \pm \sqrt{\varDelta} 
\end{equation}
given by the characteristic equation $z^2 - \text{tr}\bs{Q} \,z + \sigma = 0$ of the original mapping matrix $\bs{Q}$.
The matrix $\bs{S}$ which diagonalizes $\bs{Q}$ (that is, up to a scalar factor, the inverse of the eigenvector matrix of $\bs{Q}$) is given by 
\begin{equation} \label{eq:S}
 \bs{S} = 
 \begin{pmatrix} \gamma_- & -q_{1\bar{1}}  \\ -\gamma_+ & q_{1\bar{1}} \end{pmatrix} ~ \text{or} ~
 \begin{pmatrix} q_{2\bar{2}} & -\gamma_+ \\ -q_{2\bar{2}} & \gamma_- \end{pmatrix}
\end{equation}
if $q_{1\bar{1}} \neq 0$ or $q_{2\bar{2}} \neq 0$, respectively, where 
\begin{equation}
 \gamma_\pm = \frac{1}{2} \left( q_{1\bar{2}} + q_{2\bar{1}} \pm \sqrt{(q_{1\bar{2}} - q_{2\bar{1}})^2 + \sigma w^2} \right),
\end{equation}
with both matrices being equivalent if both $q_{1\bar{1}}, q_{2\bar{2}} \neq 0$.
$\bs{S}$ trivially equals the unit matrix if $q_{1\bar{1}} = q_{2\bar{2}} = 0$ (that is, if $\bs{Q}$ is already diagonal).
Recall here that the trace $\text{tr}\bs{Q} = 2i(q_{1\bar{2}} - q_{2\bar{1}})/w = z_+ + z_-$, the determinant $\sigma = z_+z_-$, and thereby also the discriminant $\varDelta = (\text{tr}\bs{Q}/2)^2 - \sigma$ in \cref{eq:eigenvalues} remain invariant under similarity transformations in the solution space and are therefore real quantities, since any complex basis $\bs{\phi}$ can be similarity-transformed into a real one.

The basis $\bs{\chi}$ is `symmetry-adapted' in $\mathcal{D}$ in the sense that the $\chi_\pm$ are eigenfunctions of the operator $\hat{O}_F$ corresponding to the symmetry transform $F$ acting in $\mathcal{D}$.
Indeed, since $\bs{Q}_\chi$ is diagonal, the LSB functions $\chi_\pm$ are not coupled upon their mapping, that is, each function is separately given by a constant factor times its image throughout the local symmetry domain $\mathcal{D}$:
\begin{equation} \label{eq:Fchi}
 \hat{O}_F\chi_\pm(x) = \chi_\pm(\bar{x}) = z_\pm \chi_\pm(x).
\end{equation}

For inversion symmetry ($\sigma = -1$) we always have $\text{tr}\bs{Q}=0$ from \cref{eq:q_mn} and hence $\varDelta = 1$, so that the mapping factors from \cref{eq:eigenvalues} are
\begin{equation}
 z_\pm = \pm 1 ~~~~~ (F = \varPi).
\end{equation}
Thus, $\chi_\pm \equiv \chi_{e,o}$ is here the LSB of even and odd solutions in the domain $\mathcal{D}$ of a locally $\varPi$-symmetric potential with respect to its inversion point $\alpha$.

For translation symmetry ($\sigma = 1$) we can write $z_+ = z_-^{-1} \equiv |z|e^{ikL}$, with the discriminant $\varDelta$ distinguishing three cases for the $z_\pm$ from \cref{eq:eigenvalues} (recall that $\text{tr}\bs{Q}$ and $\varDelta$ are real, as explained above):

(a) If $\varDelta < 0$, then the $\chi_\pm \equiv \chi_{\pm k}$ are propagating wave solutions with complex conjugate mapping factors 
\begin{equation} \label{eq:mapping_factors} 
 z_\pm = e^{\pm ikL} ~~~~~ (F = T)
\end{equation}
under translation by $L$ with $kL = \arctan(2\sqrt{-\varDelta}/\text{tr}\bs{Q})$, where $|z_\pm|^2 = \sigma = 1$ is accordance with the conservation of each current $j_\pm$.
Equation (\ref{eq:loc_bas_bound_cond}) for the $\chi_\pm$ coincides with the condition corresponding to global potential symmetry for the pure mapping relation (\ref{eq:pure_mapping}), that is, with vanishing one-function $q$, as shown in Ref.\,\cite{Kalozoumis2014_PRL_113_050403_InvariantsBroken}.
Therefore, $k$ is identified as the crystal momentum in the corresponding Bloch state (at energy $E$) for $\mathcal{D} = \mathbb{R}$.

(b) If $\varDelta > 0$, then the mapping factors can be written as real exponentials $z_\pm = e^{\pm \kappa L}$ (i.\,e., in the form of \cref{eq:mapping_factors} for imaginary $k \equiv -i\kappa$) since $z_+z_- = \sigma = 1$, where $\kappa L = \ln(\text{tr}\bs{Q}/2 + \sqrt{\varDelta})$.
The associated solutions $\chi_\pm \equiv \chi_{\pm \kappa}$ must now be real (up to constant phase factors) in order to conserve zero current under translation.
Lying energetically in the gaps between allowed energy bands for the corresponding globally periodic system (with symmetry domain $\mathcal{D} = \mathbb{R}$), these solutions diverge at $\pm\infty$ for $\kappa > 0$, and can be involved in physically acceptable solutions only for setups with finite (or semi-infinite) locally symmetric domains.
Enhancing the contribution of components of type $\chi_{+\kappa}$ ($\chi_{-\kappa}$) in a physical---propagating or not---state on the left (right) of a boundary between local $T$-symmetry domains may then enable controllable density accumulation, that is, wave localization, around the boundary.

(c) If $\varDelta = 0$, then \cref{eq:eigenvalues} exhibits a double root $z_\pm = 1$ or $-1$ corresponding to $k=0$ or to $k = \pm\pi/L$ (modulo $2\pi$) in \cref{eq:mapping_factors}, with associated solutions which are periodic with period $L$ or $2L$, respectively. 

\section{Construction of global basis and physical solution}
\label{sec:physical_solution}

The local basis approach developed above can be used to construct a global solution basis $\bs{\xi}(x)$ for a potential with arbitrary combinations of $\varPi$- or $T$-symmetry domains by matching of the different LSB solutions at the domain interfaces, since $\bs{\xi}(x)$ need be continuous (and smooth for nondiverging potentials).
The procedure enables an efficient assembly of this global basis $\bs{\xi}(x)$ which exploits the local symmetries of the system through the mapping of the LSBs via $\bs{Q}$-matrices along each symmetry domain. 
While the LSB mapping was formulated above for a single domain $\mathcal{D}$, we will now consider multiple attached domains and LSB mappings between consecutive cells within each domain.
A labeling for domains and cells is thus introduced as follows:
We consider a spatial decomposition of a given potential into $N$ domains $\mathcal{D}_\dom = [x_{\dom-1},x_\dom ]$ ($\dom = 1:N \equiv 1,2,...,N$) which obey distinct symmetry transformations $F_d$ (that is, are characterized by different inversion centers $\alpha_\dom$ or periods $L_\dom$).
A domain $\mathcal{D}_\dom$ is further divided into $N_\dom$ cells $\mathcal{C}^{(\dom)}_{\cell=1:N_\dom}$ of equal length (see \cref{fig:ls_domains}).
In a $T$-symmetry domain each cell covers a period $L_\dom$, while a $\varPi$-symmetry domain is divided into a left and a right cell by its inversion point $\alpha_d$ (so that always $N_\dom=2$ for $\varPi$-symmetry).
Since the cell index has no indication of the domain it belongs to, each $l$-subscripted object is also $(d)$-superscripted as in $\mathcal{C}^{(\dom)}_{\cell}$.

We now apply the LSB mapping relation, \cref{eq:chi_mapping,eq:Q_LSbasis} (or, equivalently, \cref{eq:Fchi}), to the cells of a given domain:
Within a domain $\mathcal{D}_d$, the LSB solution $\bs{\chi}^{(\dom)}$ (obtained from an arbitrary basis $\bs{\phi}^{(\dom)}$ through \cref{eq:S_transform}) is propagated from cell to cell by a diagonal mapping matrix $\bs{Q}^{(\dom)}_\chi$, so that
\begin{equation} \label{eq:cell_propagation}
 \bs{\chi}^{(\dom)}_l(F_d^{l-1}(x)) = [\bs{Q}^{(\dom)}_\chi]^{l-1} \bs{\chi}^{(\dom)}_1(x), ~~ x \in \mathcal{C}^{(\dom)}_1
\end{equation}
gives the LSB amplitude profile $\bs{\chi}^{(\dom)}_l(x)$ in the $l$-th cell of $\mathcal{D}_d$ in terms of the one in its first cell through the symmetry transform $F_d$ of the domain acting $l-1$ times.

Note here that the invariance (constancy) of the $\bs{Q}^{(d)}$-matrix of the initial solution vector $\bs{\phi}^{(d)}$ within a locally symmetric domain $\mathcal{D}_\dom$ allows for its diagonalization into $\bs{Q}^{(\dom)}_\chi$ at \textit{any} pair of symmetry-related points in $\mathcal{D}_\dom$.
Convenient points, requiring knowledge of $\bs{\phi}^{(d)}$ in minimally extended regions, are the inversion point $\alpha_\dom$ and the endpoints $x_\dom,x_\dom+L$ of the first period for $\varPi$- and $T$-symmetry, respectively.

We proceed assuming that we have found the different LSB functions of all $N$ domains, which will now be connected by matching them at the domain interfaces.
As mentioned above, the matching is necessary to construct a continuous (and smooth) global basis on which a physical solution can be represented later.
The LSB in $\mathcal{D}_\dom$ will generally match a linear combination of the LSB solutions $\bs{\chi}^{(\dom+1)}$ of the next domain $\mathcal{D}_{\dom+1}$ at the interface $x_{\dom}$,
\begin{equation} \label{eq:matching}
 \bs{\chi}^{(\dom)}(x_{\dom}) = \bs{M}_{\dom+1} \bs{\chi}^{(\dom+1)}(x_{\dom}),
\end{equation}
with a matching matrix $\bs{M}_{\dom+1}$.
Note that the interface point $x_{\dom}$ may be considered to belong to the domain $\mathcal{D}_\dom$ or to $\mathcal{D}_{\dom+1}$, to both, or even to neither of them, depending on whether the symmetry transform in $\mathcal{D}_\dom$ and/or $\mathcal{D}_{\dom+1}$ applies for this boundary point or not. 
In any case, though, its left neighborhood $x_{\dom}^-$ belongs to $\mathcal{D}_\dom$ and its right neighborhood $x_{\dom}^+$ belongs to $\mathcal{D}_{\dom+1}$. 
Therefore, the matching conditions can be expressed in a general way using $\bs{\chi}^{(\dom)}(x_{\dom}^-)$ and $\bs{\chi}^{(\dom+1)}(x_{\dom}^+)$ in \cref{eq:matching}.  
Special care is needed to handle the case when the potential is singular at $x_{\dom}$ (containing, e.\,g., terms proportional to $\delta(x-x_d)$), whence the matching conditions (and thus the matching matrix) should be adapted accordingly to the discontinuity of the wavefunction derivative.
For a potential which allows for a continuous wavefunction derivative at $x_{\dom}$, the general matching matrix reads
\begin{equation}
 \bs{M}_{\dom+1} = \frac{1}{W^{d+1,d+1}_{+,-}} 
 \begin{pmatrix} 
 W^{\dom,\dom+1}_{+,-}  & W^{\dom+1,\dom}_{+,+} \\ W^{\dom,\dom+1}_{-,-} & W^{\dom+1,\dom}_{+,-} 
 \end{pmatrix}
\end{equation}
with shorthand notation 
\begin{equation}
 W^{i,j}_{r,s} \equiv [ \chi^{(i)}_r \chi^{(j)\prime}_s - \chi^{(j)}_s \chi^{(i)\prime}_r]_{x=x_{d}},
\end{equation}
following from the continuity of the functions $\bs{\chi}^{(\dom)}(x),\bs{M}_{\dom+1} \bs{\chi}^{(\dom+1)}(x)$ and their first derivatives at $x=x_{\dom}$.

The aim is now to combine the above procedures---diagonal propagation of each domain's LSB among its cells, \cref{eq:cell_propagation}, and matching of different LSBs at domain interfaces, \cref{eq:matching}---to obtain a continuous basis for the whole potential region, denoted $\bs{\xi}(x)$, which consists of connected parts $\bs{\xi}^{(\dom)}(x)$ (with $x \in \mathcal{D}_\dom$) corresponding to the different domains.
Specifically, we start from an initial ($i$) desired reference domain $\mathcal{D}_i$, for which we set 
\begin{equation} \label{eq:xi_initial}
 \bs{\xi}^{(i)}(x) \equiv \bs{\chi}^{(i)}(x), ~~~ x \in \mathcal{D}_i
\end{equation}
from which we shall construct the global basis $\bs{\xi}(x)$ by applying matching conditions at the consecutive domain interfaces.
Indeed, according to \cref{eq:matching}, at another domain $\mathcal{D}_{d}$ with $d>i$ the corresponding part $\bs{\xi}^{(\dom)}(x)$ of the global basis will equal the LSB of that domain multiplied by the product $\bs{M}^{(\dom i)}$ of consecutive matching matrices from $i$ to $d$:
\begin{align}
 \bs{\xi}^{(\dom)}(x) = \left\{ \prod_{\dom'=i+1}^{\dom} \bs{M}_{\dom'} \right\} \bs{\chi}^{(\dom)}(x) 
 \equiv \bs{M}^{(\dom i)} &\bs{\chi}^{(\dom)}(x), ~~~ x \in \mathcal{D}_d.
\end{align}
However, each LSB $\bs{\chi}^{(\dom)}$ in the $l$-th cell of the corresponding domain $\mathcal{D}_\dom$ can be obtained from the first cell through \cref{eq:cell_propagation}, with its argument back-transformed by the inverse transform $F_d^{-1}$ applied $l-1$ times,
\begin{equation} \label{eq:cell_backpropagation}
 \bs{\chi}^{(\dom)}_l(x) = [\bs{Q}^{(\dom)}_\chi]^{l-1} \bs{\chi}^{(\dom)}_1(F_d^{-(l-1)}(x)), ~~ x \in \mathcal{C}^{(\dom)}_l
\end{equation}
for all cells $l=1:N_d$.
Thus, on the level of cells, the global basis can be written as a branched function
\begin{equation} \label{eq:xi_branch}
 \bs{\xi}(x) 
 = \bs{\xi}^{(\dom)}_l(x) = \bs{G}^{(\dom i)}_l \bs{\chi}^{(\dom)}_1(F_d^{1-l}(x)), ~~ x \in \mathcal{C}^{(\dom)}_l,
\end{equation}
where the (forward) basis propagation-matching matrix 
\begin{equation} \label{eq:lsb_transfer_matrix}
 \bs{G}^{(\dom i)}_l  =  \bs{M}^{(\dom i)} [\bs{Q}^{(\dom)}_\chi]^{\cell-1}
\end{equation}
first propagates $\bs{\chi}^{(\dom)}$ from the first to the $l$-th cell in $\mathcal{D}_\dom$ and then applies the matching up to this domain.
The multidomain basis $\bs{\xi}(x)$ is determined by the LSB in the initial domain $\dom = i$ in the sense of \cref{eq:xi_initial}.
The imprint of the local symmetry of the potential is manifest in $\bs{\xi}$ through the mapping
\begin{equation}
 \bs{\xi}^{(\dom)}(\bar{x}) = \bs{Q}_\xi^{(\dom)}  \bs{\xi}^{(\dom)}(x), ~~ x \in \mathcal{D}_d
\end{equation}
within each domain via the corresponding (transformed) constant mapping matrix 
\begin{equation}
  \bs{Q}_\xi^{(\dom)} =  \bs{M}^{(\dom i)} \bs{Q}_\chi^{(\dom)} [\bs{M}^{(\dom i)}]^{-1}.
\end{equation}
This reveals a universal structure of the solution space for potentials with local symmetries in terms of domainwise invariants.

For $f < i$, the (backward) basis propagation-matching (from cell $N_\dom$ to $l$ and from domain $i$ to $\dom < i$) is performed by the matrix
\begin{equation}
 \tilde{\bs{G}}^{(i \dom)}_l = [\bs{M}^{(i \dom)}]^{-1}[\bs{Q}^{(\dom)}_\chi]^{-N_\dom + \cell}
\end{equation}
containing $\dom-i$ matching matrix inversions, with the diagonal $\bs{Q}^{(\dom)}_\chi$-matrices elementwise inverted.

If $i=1$ and $f=N$, then $\bs{\xi}(x)$ constitutes a global basis for the complete potential region, on which the physical solution $\psi$ is expanded as
\begin{equation} \label{eq:psi_global}
\psi(x) = \bs{c} \cdot \bs{\xi}(x) = c_1 \xi_1(x) + c_2 \xi_2(x),
\end{equation}
with the amplitude vector $\bs{c}$ determined by the boundary conditions imposed at $x=x_0,x_N$.
As indicated above, the role of $\bs{G}^{(\dom i)}_l$ is to propagate the LSB function $\bs{\chi}^{(\dom)}$ from the first to the $l$-th cell in $\mathcal{D}_\dom$ and subsequently apply the domain interface matching up to this domain (see \cref{eq:xi_branch}) given an initial domain $i$ with $\bs{\xi}^{(i)} = \bs{\chi}^{(i)}$ (see \cref{eq:xi_initial}).
Thus, if the physical solution has local coefficients $\bs{a} \equiv \bs{a}^{(\dom)}_l$ in an arbitrary basis $\bs{\phi}^{(\dom)}$ (see end of \cref{sec:mixed_mapping}) in the $l$-th cell of domain $\mathcal{D}_\dom$, then those are related to the coefficients $\bs{c}$ in the constructed global basis $\bs{\xi}$ as $\bs{a}^{(\dom)}_l = \bs{c} \, \bs{G}^{(\dom i)}_l \bs{S}^{(\dom)}$, where  $\bs{S}^{(\dom)}$ is the matrix transforming $\bs{\phi}^{(\dom)}$ to $\bs{\chi}^{(\dom)}$ in domain $\mathcal{D}_\dom$ (see \cref{eq:S_transform}).
The application of boundary conditions is thus naturally postponed until a basis of the solution space for the complete potential has been obtained, offering flexibility with respect to the setup at hand:
Energy-quantizing (e.\,g. Dirichlet, Neumann, mixed, periodic, or exponentially decaying) boundary conditions determine $c_1$ and $c_2$ subject to appropriate normalization, while continuous-spectrum (scattering) asymptotic conditions relate $c_{1,2}$ to propagating wave amplitudes at both ends (see discussion below).

Note here the conceptual difference of the local basis approach from a conventional transfer matrix method where the amplitude vectors of $\psi$ are propagated in a fixed basis (usually of counterpropagating plane waves in flat potential regions):
Here, instead of the physical solution, the basis itself is propagated in a locally symmetric setup with (repeated or inverted) unit cells of arbitrary potential profile.
Even in cases of, e.\,g., intervals of finite periodic potentials, the corresponding basis in the interval is usually adapted from the globally periodic counterpart \cite{Stslicka2002_SurfSciRep_47_93_LocalisedElectronicStatesInSCSuperlattices} with explicit spatial dependence.
Here, the LSB is constructed \textit{intrinsically} from an arbitrary solution of the first cell of the local symmetry domain.
In particular, the present approach exploits the local symmetries by virtue of the symmetry-adapted bases which are propagated (forward or backward) through multiple cells by diagonal $\bs{Q}$-matrices, thus providing an important technical advantage---especially in the presence of large periodic parts.

It should be pointed out that, although the present approach is devised for potentials which are decomposable into multiple local symmetry domains, its application does not become invalid in presence of \textit{nonsymmetric} domains (such as, e.\,g., defects in a finite periodic lattice).
Specifically, we can simply treat a domain $\mathcal{D}_\dom$ which is neither $\varPi$- nor $T$-symmetric (such as $\mathcal{D}_1$ or $\mathcal{D}_4$ in \cref{fig:ls_domains}) as a domain with a single cell $\mathcal{C}^{(\dom)}_{\cell=1}$ in the used notation, and use \cref{eq:xi_branch,eq:lsb_transfer_matrix} with $l=1$ for this domain (coinciding with the cell).
It is then clear that no $\bs{Q}^{(\dom)}_\chi$ is involved for this nonsymmetric domain, which makes sense since there is no local symmetry to be exploited;
only the matching matrix $\bs{M}^{(\dom i)}$ remains in \cref{eq:lsb_transfer_matrix} to match this domain's local basis $\bs{\chi}^{(\dom)}_{\cell=1}$ to that of the previous domain.
Further, since there is no symmetry to adapt the basis to in the domain, the basis $\bs{\chi}^{(\dom)}$ can be chosen arbitrarily.
Therefore, it is simply set equal to the initially computed basis, $\bs{\chi}^{(\dom)} = \bs{\phi}^{(\dom)}$, for (the first and only cell of) this nonsymmetric domain. 
In other words, the construction of the global basis $\bs{\xi}$ can still be applied if some domains of the setup happen to be nonsymmetric, although clearly no symmetry-induced advantage can be drawn from these domains.

Let us now summarize the procedure followed in the LSB approach to stationary wave systems, as schematically represented by the sequence
\begin{equation} \label{eq:sequence}
 \bs{\phi}^{(\dom)}(\mathcal{C}^{(\dom)}_1) ~~
 \xrightarrow[~\alpha_\dom,L_\dom~]{\bs{Q}^{(\dom)}} ~~
 z^{(\dom)}_\pm,\bs{S}^{(\dom)} 
 \xrightarrow[~~~~~]{} ~~
 \bs{\chi}^{(\dom)}_1 ~~ \forall \, \mathcal{D}_\dom ~~
 \xrightarrow[~~~~~]{\bs{G}} ~~
 \bs{\xi}(x) ~~
 \xrightarrow[~~~~~]{\bs{c}} ~~ 
 \psi 
\end{equation}
and expressed as follows:
\begin{itemize}
 \item[(i)] Decompose the potential into $N$ domains $\mathcal{D}_{\dom=1:N}$ containing maximal regions of local $\varPi$- or $T$-symmetry, and compute an arbitrary pair of linearly independent solutions $\bs{\phi}^{(\dom)}(x)$ (if possible analytically, or numerically with arbitrary initial conditions) to \cref{eq:schrodinger} only in the first cell $\mathcal{C}^{(\dom)}_1$ of each domain. 
 \item[(ii)] Construct the matrix $\bs{Q}^{(\dom)}$ from $\bs{\phi}^{(\dom)}(x)$ at $x=\alpha_\dom$ ($x_\dom,x_\dom+L$) for local $\varPi$- ($T$-) symmetry, and diagonalize it to find its eigenvalues $z^{(\dom)}_\pm$ and $\bs{S}_\dom$-matrix in the basis $\bs{\phi}^{(\dom)}$.
 \item[(iii)] Propagate and transform the first-cell LSBs $\bs{\chi}^{(\dom)}_1 = \bs{S}_\dom \bs{\phi}^{(\dom)}(x \in \mathcal{C}^{(\dom)}_1)$ within each domain $\mathcal{D}_\dom$ by the matrices $\bs{G}^{(\dom i)}$, $\tilde{\bs{G}}^{(i \dom)}$ with a selected initial (reference) domain $i$ and final (end) domains $f=1,N$ to obtain a global basis $\bs{\xi}(x)$ in the potential region.
\item[(iv)] Impose desired boundary conditions on a physical solution $\psi(x) = \bs{c} \cdot \bs{\xi}(x)$.
\end{itemize}

Recall that the global basis $\bs{\xi}$ in step (iii) coincides with the local basis $\bs{\chi}^{(i)}$ in the selected domain $\mathcal{D}_i$ which can be anywhere in the interaction region.
Assuming that the potential can be, to some extent, tuned by external parameters, one could design a desired (for simplicity, nodeless) wave profile for $\bs{\chi}^{(i)}$ and determine the corresponding---not necessarily locally symmetric---domain potential $V_i$ from \cref{eq:schrodinger} as \cite{Cooper1995_PR_251_267_SupersymmetryQuantum} $V_i(x) = E + \chi_\pm^{(i)\prime\prime}(x)/2\chi_\pm^{(i)}(x)$.
Given the LSB mapping within and among domains, and in particular solutions with exponential mapping factors $z_\pm = e^{\mp\kappa L}$, this provides enhanced controllability of the spatial field distribution such as its localization in selected regions:
While the coefficients $c_{1,2}$ are uniquely determined in the case of energy-quantizing boundary conditions, in the case of scattering we can impose $c_{1,2}=1$ and solve for the amplitudes $a_\pm^{<,>}$ of plane waves $e^{\pm i\sqrt{2E}x}$ on the left ($x<x_0$) and on the right ($x>x_N$) of the interaction region.
In other words, the ingoing amplitudes $a^<_+$ and $a^>_-$ that produce a desired domain localization at a given energy in a locally symmetric potential can be determined efficiently by the proposed scheme.

\section{Summary and conclusion}\label{sec:conclusion}

Employing the two-dimensional solution space of the stationary Schr\"odinger equation, we have developed a theory for treating 1D wave mechanical systems with local (i.\,e. domainwise) inversion and translation symmetries.
Encoding the local symmetries of the potential into corresponding local basis vectors, the formalism is independent of the boundary conditions imposed subsequently on particular physical solutions.
The approach is based on two-function, symmetry-induced local invariants, which have the form of two-point currents and are spatially constant within each domain of local symmetry.
They enable an extension of the generalized 1D parity and Bloch theorems of Ref.\,\cite{Kalozoumis2014_PRL_113_050403_InvariantsBroken}, i.\,e. domainwise amplitude mapping relations, to arbitrary boundary conditions and thereby to states carrying zero current.
More importantly, the theory reveals a universal structure of wavefunctions in locally symmetric potentials in terms of special local bases which are adapted to a given symmetry in a finite domain. 
The local symmetry bases (LSBs) are constructed from arbitrary initial solutions in only a single unit cell (one half of an inversion symmetry domain or one period of a translation symmetry) and mapped among cells by diagonal matrices.
Combined with the matching of different LSBs at symmetry domain interfaces, this leads to an efficient computational scheme for spatially resolved wavefunctions in systems with arbitrary combinations of local inversion and translation symmetries.
The advantage of the method is especially pronounced for completely locally symmetric (CLS) systems with different large periodic parts.
The multiplicative mapping of local basis functions within each locally periodic domain by exponentials then enables a natural control of wave amplitude distribution by tuning the potential parameters.
In particular, input amplitudes in scattering setups can be flexibly designed to produce localization in desired spatial domains.
Valid generically for wave mechanics (e.\,g. acoustics, optics, or quantum mechanics), the LSB approach provides the perspective to use local symmetries to explain and control the amplitude response of structurally complex scattering or bound systems.

\section*{Acknowledgments}

P.A.K acknowledges financial support from IKY Fellowships of Excellence for Postdoctoral Research in Greece - Siemens Program.

\appendix

\begin{figure}
\includegraphics[width=1\columnwidth]{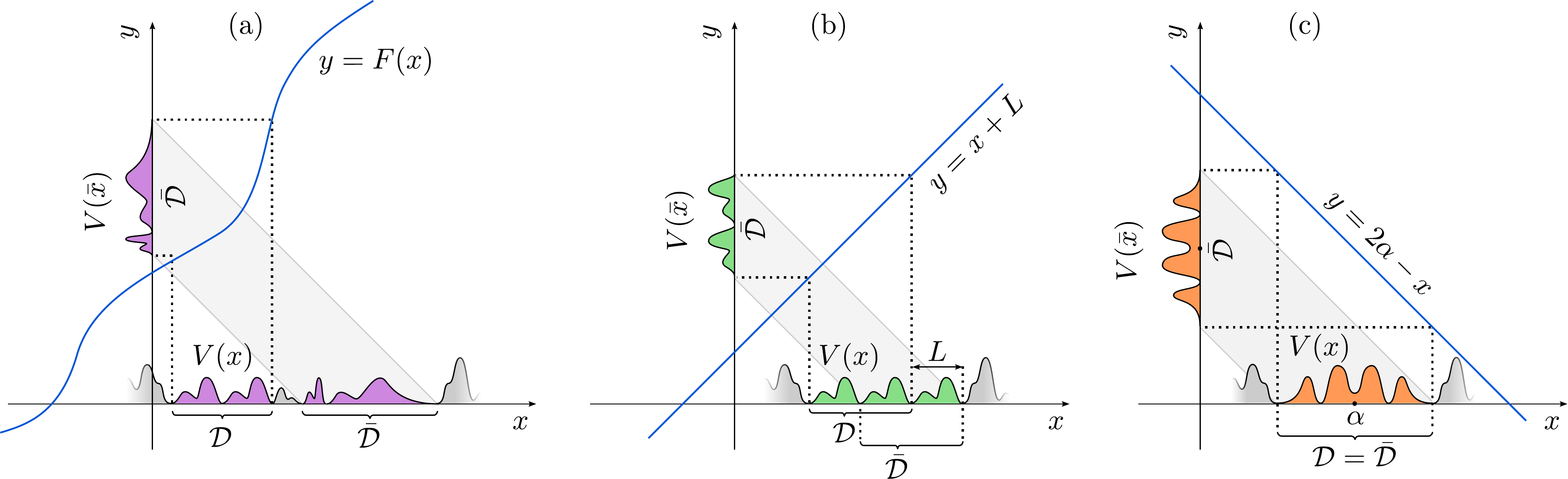}
\caption{ 
Local symmetry of a potential $V(x)$ within a domain $\mathcal{D}$ under a similarity transformation $F:x \rightarrow y = F(x)$ (blue lines) which maps $\mathcal{D}$ to $\bar{\mathcal{D}}$ with potential $V(\bar{x})$ for (a) a general bijective transformation $F$, (b) translation by $L$, $y = T(x) = x + L$, and (c) inversion through $\alpha$, $y = \varPi(x) = 2\alpha - x$. The locally symmetric part of the potential (colored) is generally embedded between nonsymmetric parts (gray). Shaded stripes highlight the imaging of the transformed potential $V(\bar{x})$ onto the $x$-axis.}
\label{fig:general_transform}
\end{figure}

\section{Local two-point invariants for general spatial transformations}

We here derive bilinear quantities analogous to the form in \cref{eq:q_mn} which are spatially constant for general symmetry transformations of the potential of a given domain and enable a generalization of the mapping relation in \cref{eq:mixed_mapping}.
Specifically, let us consider an arbitrary bijective coordinate transformation $F:x \rightarrow y=F(x)$ and a domain $\mathcal{D}$ mapped through $F$ to a domain $\bar{\mathcal{D}}$, for which the potential obeys $V(x) = V(y)$ with $x \in \mathcal{D}$ and $y \equiv \bar{x} \in \bar{\mathcal{D}}$, as shown in \cref{fig:general_transform}\,(a).
The aim is to construct a two-function quantity $Q_F(x,y)$ whose total derivative with respect to $x$, in analogy to \cref{eq:mixed_mapping}, vanishes under the above $F$-symmetry (or `shape invariance') of the potential.
With $Q_F$ being a function of $y$, its derivative will generally be affected by the transformation $F$.
To exploit the local $F$-symmetry of the potential, we therefore write the Schr\"odinger equation in the transformed coordinate $y$, which becomes
\begin{equation}
\label{eq:schrodinger_transformed} 
H_F\psi_F(y) \equiv -\frac{1}{2} D_F(y) + V(y) \psi_F(y) = E \psi_F(y)
\end{equation}
with the second derivative transformed to
\begin{equation}
\label{eq:D_F} 
D_F(y) = F''(F^{-1}(y)) \,\dot{\psi}_F(y) + [F'(F^{-1}(y))]^2 \,\ddot{\psi}_F(y)
\end{equation}
using the chain rule $\frac{d}{dx}\psi_F(y(x)) = F'(x)\frac{d}{dy} \psi_F(y)$ with $x = F^{-1}(y)$ ($F^{-1}$ being the inverse coordinate transform), where we define $F'(s) = \left.\frac{dF(x)}{dx}\right|_{x=s}$ and $\dot{\psi}_F = \frac{d}{dy} \psi_F$. 
The function $\psi_F(y)$ denotes a solution of the modified equation arising from the deformation of the single axis of our 1D system.
Under the local shape invariance of the potential, the bilinear two-point combination
\begin{equation}
\label{eq:Qgen} 
Q_F(x,y) = \frac{1}{2i} \left[\psi(x)F'(x) \dot{\psi}_F(y) - \psi'(x)\psi_F(y) \right]
\end{equation}
of a solution $\psi(x)$ of \cref{eq:schrodinger} and a solution $\psi_F(y)$ of \cref{eq:schrodinger_transformed} is then spatially constant in $\mathcal{D}$, since
\begin{equation} \label{eq:total_derivative_F}
2iQ_F' = \psi(x)D_F(y) - \psi''(x)\psi_F(y) = 0
\end{equation}
from \cref{eq:schrodinger,eq:schrodinger_transformed} for $V(x) = V(y)$ with $x \in \mathcal{D}$, $y \equiv \bar{x} \in \bar{\mathcal{D}}$.
This invariant quantity $Q_F$ is the generalized version of $Q$ in \cref{eq:Qpure} for a general (smooth) similarity transform $F$, with alternative $\tilde{Q}_F$ defined in the same manner (replacing $\psi$ by $\psi^*$ in \cref{eq:Qgen}).

In particular, we can express \cref{eq:Qgen} for basis functions $\phi_1(x),\phi_2(x)$ and $\phi^F_1(y),\phi^F_2(y)$ of the solution spaces of \cref{eq:schrodinger,eq:schrodinger_transformed}, respectively, as 
\begin{equation} \label{eq:q_mn_gen} 
q^F_{m\bar{n}} = \frac{1}{2i}  \left[\phi_m(x)F'(x) \dot{\phi}^F_n(y) - \phi_m'(x)\phi^F_n(y) \right],
\end{equation}
in analogy with \cref{eq:q_mn}, with $q^F_{m\bar{n}}(x,\bar{x})$ being spatially constant in $\mathcal{D}$.
These generalized invariants now map the local basis $\bs{\phi}(x)$ in the original axis within a domain $\mathcal{D} \ni x$ to the basis $\bs{\phi}_F(\bar{x})$ in the transformed axis within the image domain $\bar{\mathcal{D}} \ni \bar{x} = y$:
\begin{equation} \label{eq:mixed_mapping_gen}
 \bs{\phi}_F(\bar{x}) = \bs{Q}_F \bs{\phi}(x),
\end{equation}
where the mapping matrix $\bs{Q}_F = \bs{Q}_F(x,\bar{x})$ has the same form as $\bs{Q}$ in \cref{eq:Q_matrix} but with the $q_{m\bar{n}}$ replaced by $q^F_{m\bar{n}}$, as simply shown in the same manner as \cref{eq:mixed_mapping} from \cref{eq:mixed_mapping_derivation}.

In the present article we focus on the symmetry-induced mapping of a \textit{single} basis between $F$-mapped domains spanning the solution space of the Schr\"odinger equation in a fixed coordinate system.
In other words, we demand that the basis functions $\phi^F_{1,2}$ in \cref{eq:q_mn_gen} (or equivalently, the function $\psi_F$ in \cref{eq:Qgen}) be solutions of the original equation (\cref{eq:schrodinger}), which is nontrivially the case only if $D_F(y) = \ddot{\psi}_F(y)$ in \cref{eq:schrodinger_transformed}.
In other words, we here consider transformations which leave the Schr\"odinger equation invariant, so that the same basis $\bs{\phi}$ (or solution $\psi$) at two points $x,\bar{x}$ can be used in \cref{eq:q_mn_gen} (or in \cref{eq:Qgen}).
Since this should hold for arbitrary locally $F$-symmetric potential (and thereby arbitrary $\bs{\phi}$ or $\psi$), we demand that (cf. \cref{eq:D_F})
\begin{equation}
\label{eq:isometry} 
F''(x) = 0, ~~[F'(x)]^2 = 1  ~~ \Rightarrow ~~ F(x) = \sigma x + \rho, ~~\sigma = \pm 1,
\end{equation}
so that $F$ is an isometry corresponding to the (local) inversion or translation transforms in \cref{eq:F}.
Further, local symmetry transforms of this type between finite domains $\mathcal{D}$ and $\bar{\mathcal{D}}$ have a global limit with $\mathcal{D}=\mathbb{R}$, which is not the case for an arbitrary (non-isometric) transformation $F$.
The two particular cases of local inversion and translation symmetry (see \cref{fig:general_transform}\,(b) and (c), respectively) entailed in \cref{eq:isometry} then enable the recovery of stationary parity and Bloch eigenfunctions for the Hamiltonian in the global symmetry limit.

\end{document}